\documentclass[10pt, final, journal]{IEEEtran} 
\usepackage{latexsym}
\usepackage{amsfonts}
\usepackage{cite}
\usepackage{amssymb}
\usepackage{bm}
\usepackage{amsmath}
\usepackage{amssymb}
\usepackage{graphicx}
\IEEEoverridecommandlockouts

\usepackage{epsfig}
\usepackage{psfrag} 
\usepackage{setspace}
\usepackage{hyperref}
\usepackage{algorithm}
\usepackage{algpseudocode}

\usepackage{theorem}
\theoremheaderfont{\bfseries\upshape} \theoremstyle{plain}

\theorembodyfont{\rmfamily}

\theoremheaderfont{\itshape}

\theorembodyfont{\itshape}
\newtheorem{theorem}{Theorem}
\newtheorem{proposition}{Proposition}
\newtheorem{lemma}{Lemma} 

\newtheorem{definition}{Definition}

\newtheorem{remark}{Remark}

\def \cX {\mathcal{X}}

\def \bR {\mathbb{R}}

\def \Q#1 {{\sc Question~#1}~}

\makeatletter 
  \newcommand\figcaption{\def\@captype{figure}\caption} 
  \newcommand\tabcaption{\def\@captype{table}\caption} 
\makeatother

\title{Information Measures: \\ the Curious Case of the Binary Alphabet}

\author{Jiantao~Jiao,~\IEEEmembership{Student Member,~IEEE},~Thomas~A. Courtade,~\IEEEmembership{Member,~IEEE},~Albert~No,~\IEEEmembership{Student Member,~IEEE},~Kartik~Venkat,~\IEEEmembership{Student Member,~IEEE}, and Tsachy~Weissman,~\IEEEmembership{Fellow,~IEEE}

\thanks{Manuscript received Month 00, 0000; revised Month 00, 0000; accepted Month 00, 0000. Date of current version Month 00, 0000.
This work was supported in part by the Center for Science of Information (CSoI), an NSF Science and Technology Center, under grant agreement CCF-0939370. The material in this paper was presented in part at the 2014 IEEE International Symposium on
Information Theory, Honolulu, HI, USA. Copyright (c) 2014 IEEE. Personal use of this material is permitted. However, permission to use this material for any other purposes must be obtained from the IEEE by sending a request to pubs-permissions@ieee.org. 
}%

\thanks{J.~Jiao, A.~No, K.~Venkat, and T.~Weissman are with the Department of Electrical Engineering, Stanford University. Email: \{jiantao, albertno, kvenkat, tsachy\}@stanford.edu}
\thanks{T. Courtade is with the Department of Electrical Engineering and Computer Science, University of California, Berkeley. Email: {courtade@eecs.berkeley.edu}}
}

\begin{document}

\maketitle
\thispagestyle{plain}
\pagestyle{plain}

\begin{abstract}
Four problems related to information divergence measures defined on finite alphabets are considered.    In three of the cases we consider, we illustrate a contrast which arises between the binary-alphabet and larger-alphabet settings.  This is surprising in some instances, since characterizations for the larger-alphabet settings do not generalize  their binary-alphabet counterparts.  Specifically, we show that $f$-divergences are not the unique decomposable divergences on binary alphabets that satisfy the data processing inequality, thereby clarifying claims that have previously appeared in the literature. We also show that KL divergence is the unique Bregman divergence which is also an $f$-divergence for any alphabet size. We show that KL divergence is the unique Bregman divergence which is invariant to statistically sufficient transformations of the data, even when non-decomposable divergences are considered. Like some of the problems we consider, this result holds only when the alphabet size is at least three. 
\end{abstract}


\section{Introduction}\label{sec.intro}
Divergence measures play a central role in information theory and other branches of mathematics.  Many special classes of divergences, such as Bregman divergences \cite{Bregman1967relaxation}, $f$-divergences \cite{Morimoto1963markov,Ali--Silvery1966general,Csiszar1967information}, and Kullback-Liebler-type $f$-distance measures \cite{Kannappan--Sahoo1992kullback}, enjoy various properties  which make them particularly useful in problems related to learning, clustering, inference, optimization, and quantization, to name a few. A review of applications of various divergence measures in statistical signal processing can be found in \cite{Basseville2013divergence}. In this paper, we  investigate the relationships between these three classes of divergences, each of which will be defined formally in due course, and the subclasses of divergences which satisfy desirable properties such as monotonicity with respect to data processing. 
 Roughly speaking, we address the following four questions:

\vspace{2ex}

\hspace{.05in}\noindent{\sc Question 1:} If a decomposable divergence  satisfies the data processing inequality, must it be an $f$-divergence?
\vspace{1ex}

\hspace{.05in}\noindent{\sc Question 2:} Is Kullback-Leibler (KL) divergence the unique KL-type $f$-distance measure which satisfies the data processing inequality?\vspace{1ex}

\hspace{.05in}\noindent{\sc Question 3:} Is KL divergence the unique Bregman divergence which is invariant to statistically sufficient transformations of the data?\vspace{1ex}

\hspace{.05in}\noindent{\sc Question 4:}  Is KL divergence the unique Bregman divergence which is also an $f$-divergence?

\vspace{2ex}

Of the above four questions, only  \Q{4}  has an affirmative answer.  However, this assertion is slightly deceiving.  Indeed, if the alphabet size $n$ is at least $3$, then all four questions can be answered in the affirmative.  Thus, counterexamples only arise in the binary setting when $n=2$. 

This is perhaps  unexpected.  Intuitively, a reasonable  measure of divergence should satisfy the data processing inequality -- a seemingly modest requirement.  In this sense, the  answers to the above series of questions imply that the class of ``interesting" divergence measures can be very small   when $n\geq 3$ (e.g., restricted to the class of $f$-divergences, or a multiple of KL divergence).  However, in the binary alphabet setting, the class of ``interesting" divergence measures is strikingly rich, a point which will be  emphasized in our results.  In many ways, this richness is surprising since the binary alphabet is usually viewed as the simplest setting one can consider.  In particular, it might be expected that the class of interesting divergence measures corresponding to binary alphabets would be less rich than the counterpart class for larger alphabets.  However, as we will see, the opposite is true.

The observation of this dichotomy between binary and larger alphabets is not without precedent.  For example, Fischer proved the following in his 1972 paper \cite{Fischer1972}.
\begin{theorem}\label{thm.fischer}
Suppose $n\geq 3$.  If, and only if, $f$ satisfies
\begin{align}
\sum_{k = 1}^n p_k f(p_k) \leq \sum_{k = 1}^n p_k f(q_k) \label{eqn.shannon}
\end{align}
for all probability distributions $P=(p_1,p_2,\ldots,p_n), Q=(q_1,q_2,\ldots,q_n)$, 
then it is of the form
\begin{equation}
f(p) = c \log p + b~~ \mbox{for all $p\in (0,1)$,} \label{FischerSolnNgeq3}
\end{equation}
where $b$ and $c\leq 0$ are constants. 
\end{theorem}
As implied by his supposition that $n\geq 3$ in Theorem \ref{thm.fischer}, Fischer observed and appreciated the distinction between the binary and larger alphabet settings when considering so-called \emph{Shannon-type} inequalities of the form \eqref{eqn.shannon}.  Indeed, in the same paper \cite{Fischer1972}, Fischer gave the following result:
\begin{theorem}\label{thm.fischern2}
The functions of the form
\begin{equation}
f(q) = \int \frac{G(q)}{q} dq, \quad q \in (0,1), \label{binSolnFischer}
\end{equation}
with $G$ arbitrary, nonpositive, and satisfying $G(1-q) = G(q)$ for $q\in(0,1)$,
are the only absolutely continuous functions\footnote{There also exist functions $f$ on $(0,1)$ which are not absolutely continuous and satisfy \eqref{eqn.shannon}.  See \cite{Fischer1972}.} on $(0,1)$ satisfying \eqref{eqn.shannon} when $n=2$. 
\end{theorem}
Only in the special case where $G$ is taken to be constant in \eqref{binSolnFischer}, do we find that $f$ is of the form \eqref{FischerSolnNgeq3}.  We direct the interested reader  to  \cite[Chap. 4]{Aczel--Darczy1975} for a detailed discussion.

In part, the present paper was inspired and motivated by  Theorems \ref{thm.fischer} and \ref{thm.fischern2}, and the distinction they draw between binary and larger alphabets.  Indeed, the answers to  \Q1 -- \Q3   are in the same spirit as Fischer's results.  For instance, our answer to  \Q2 demonstrates that the functional inequality \eqref{eqn.shannon} \emph{and} a data processing requirement are still not enough to demand $f$ take the form \eqref{FischerSolnNgeq3} when $n=2$.  To wit, we prove an analog of Theorem \ref{thm.fischern2} for this setting (see Section \ref{sec:Q2Answer}).

In order to obtain a complete picture of various questions associated with divergence measures, we also attempt to sharpen results in the literature regarding divergence measures on alphabets of at least size $3$. The most technically challenging result we obtain in this work, is to show that KL divergence is the unique Bregman divergence which is invariant to statistically sufficient transformations of the data when the alphabet size is at least three, \emph{without} requiring it to be decomposable. Indeed, dropping the decomposability conditions makes the problem much harder, and we have to borrow profound techniques from convex analysis and functional equations to fully solve it. This result is presented in Theorem~\ref{thm.bregkl}.

\subsection*{Organization}  This paper is organized as follows.  In Section \ref{sec:introToDivergence}, we recall several important classes of divergence measures and define what it means for a divergence measure to satisfy any  of the data processing, sufficiency, or decomposability properties.  In Section \ref{sec:mainResults}, we investigate each of the questions posed above and state our main results.  Section \ref{sec:Conclusion} delivers our concluding remarks, and the Appendices contain all proofs.

\section{Preliminaries: Divergences, Data Processing,  and Sufficiency Properties} \label{sec:introToDivergence}
Let $\bar{\bR} \triangleq [-\infty,+\infty]$ denote the extended real line.  Throughout this paper, we only consider finite alphabets. To this end, let $\cX = \{1,2,\ldots,n\}$ denote the alphabet, which is of size $n$, and let $\Gamma_n = \{(p_1,p_2,\ldots,p_n):\sum_{i = 1}^n p_i = 1,  p_i \geq 0, i = 1,2,\ldots,n\}$ be the set of probability measures on $\cX$, with $\Gamma_n^+ = \{(p_1,p_2,\ldots,p_n):\sum_{i = 1}^n p_i = 1,  p_i > 0, i = 1,2,\ldots,n\}$ denoting its relative interior. 

We refer to a non-negative function $D: \Gamma_n \times \Gamma_n \to \bar{\bR}_+$ simply as a divergence function (or, divergence measure).  Of course, essentially all common divergences -- including Bregman  and $f$-divergences, which are defined shortly -- fall into this general class.  In this paper, we will primarily be interested in divergence measures which satisfy either of two properties: the \emph{data processing property} or the \emph{sufficiency property}.

In the course of defining these properties, we will consider (possibly stochastic) transformations $P_{Y|X} : X \mapsto Y$, where $Y\in \cX$.  That is, $P_{Y|X}$ is a Markov kernel with source and target both equal to $\mathcal{X}$ (equipped with the discrete $\sigma$-algebra).  If $X\sim P_X$, we will write $P_X \to P_{Y|X} \to P_Y$ to denote that $P_Y$ is the marginal distribution of $Y$ generated by passing $X$ through the channel $P_{Y|X}$.  That is,   $P_{Y}(\cdot) \triangleq \sum_{x\in \cX} P_X(x) P_{Y|X}(\cdot |x)$.

Now, we are in a position to formally define the \emph{data processing property} (a  familiar concept to most information theorists).

\begin{definition}[Data Processing]\label{def.dpi}
A divergence function $D$ satisfies the data processing property if, for all $P_X,Q_X \in \Gamma_n$, we have
\begin{align}
D\left(P_X; Q_X\right) \geq D\left(P_{Y}; Q_{Y}\right),
\end{align}
for any transformation $P_{Y|X} : X \mapsto Y$, where $P_Y$ and $Q_Y$ are defined via $P_X \to P_{Y|X} \to P_Y$ and $Q_X \to P_{Y|X} \to Q_Y$, respectively. 
\end{definition}

A  weaker version of the data processing inequality is the \emph{sufficiency property}. In order to describe the sufficiency property, for two arbitrary distributions $P,Q$, we define a joint distribution $P_{XZ}$, $Z \in \{1,2\}$, such that
\begin{equation}
P_{X|1} = P,\quad  P_{X|2} = Q.
\end{equation}
A transformation $P_{Y|X} : X \mapsto Y$ is said to be a \emph{sufficient transformation of $X$ for $Z$} if $Y$ is a sufficient statistic of $X$ for $Z$.  We remind the reader that $Y$ is a sufficient statistic of $X$ for $Z$ if the following two Markov chains hold:
\begin{align}
&Z - X - Y & Z -  Y - X.
\end{align}

\begin{definition}[Sufficiency]\label{def.suff}
A divergence function $D$ satisfies the sufficiency property if, for all $P_{X|1},P_{X|2} \in \Gamma_n$ and $Z\in\{1,2\}$, we have
\begin{align}
D(P_{X|1}; P_{X|2}) \geq D\left(P_{Y|1}; P_{Y|2}\right), \label{sufficiencyDefnInequality}
\end{align}
for any sufficient transformation $P_{Y|X} : X \mapsto Y$ of $X$ for $Z$,
where $P_{Y|z}$ is defined by $P_{X|z} \to P_{Y|X} \to P_{Y|z}$ for $z\in\{1,2\}$. 
\end{definition}

We remark that our definition of {\sc sufficiency} is a variation on that given in \cite{Harremoes--Tishby2007}. 
Clearly, the sufficiency property is weaker than the data processing property because our attention is restricted to only those (possibly stochastic) transformations $P_{Y|X}$ for which $Y$ is a sufficient statistic of $X$ for $Z$.  Given the definition of a sufficient statistic, we note that the inequality in \eqref{sufficiencyDefnInequality} can be replaced with equality to yield an equivalent definition.

Henceforth, we will simply say that a divergence function $D(\cdot;\cdot)$ satisfies {\sc Data Processing}  when it satisfies the data processing property.  Similarly, we say that a divergence function $D(\cdot;\cdot)$ satisfies  {\sc Sufficiency} when it satisfies the sufficiency property.

\begin{remark}
In defining the data processing and sufficiency properties, we have required that $Y \in \cX$.  This is necessary because the divergence function $D(\cdot;\cdot)$ is only defined on $\Gamma_n \times \Gamma_n$.  
\end{remark}

Before proceeding, we make one more definition following Pardo and Vajda \cite{Pardo--Vajda1997}.
\begin{definition}[Decomposibility]
A divergence function $D$ is said to be decomposable if there exists a bivariate function $\delta(u,v): [0,1]^2 \to \bar{\bR}$ such that 
\begin{align}
D(P ; Q) = \sum_{i = 1}^n \delta(p_i, q_i)
\end{align}
for all $P=(p_1,\dots,p_n)$ and $Q=(q_1,\dots,q_n)$ in $\Gamma_n$.
\end{definition}

Having defined divergences in general, we will now recall  three important classes of divergences which will be of interest to us: Bregman divergences, $f$-divergences, and KL-type divergences.

\subsection{Bregman Divergences}

Let $G(P): \Gamma_n \to \bR$ be a convex function defined on $\Gamma_n$,  differentiable on $\Gamma_n^+$. For two probability measures $P=(p_1,\dots,p_n)$ and $Q=(q_1,\dots,q_n)$ in $\Gamma_n$, the Bregman divergence generated by $G$  is defined by
\begin{align}
D^G(P;Q) \triangleq G(P) - G(Q) -\langle \nabla G(Q) , P - Q \rangle,
\end{align}
where $\langle \nabla G(Q) , P - Q \rangle$ denotes the standard inner product between $\nabla G(Q)$ and $(P-Q)$ (interpreted as vectors in $\bR^n$).  Note that we only define Bregman divergences on the space of probability distributions in accordance with the ``information measures" theme.  It is also common to define Bregman divergences on other domains.

By Jensen's inequality, many properties of $D(P;Q)$ are readily verified.  For instance, $D^G(P;Q)\geq 0$, and hence Bregman divergences are consistent with the general definition given at the beginning of this section.

In addition to the elementary observation that $D^G(P;Q)\geq 0$, Bregman divergences enjoy a number of other properties which make them useful for many learning, clustering, inference, and quantization problems. We refer the interested reader to the recent survey \cite{Basseville2013divergence} and references therein for an overview. 

Note that  if the  convex function $G(P)$ which generates $D^G(P;Q)$ takes the form:
\begin{equation}
G(P) = \sum_{i = 1}^n g(p_i), 
\end{equation}
then $D^G(P;Q)$ is decomposable since
\begin{align}
D^G(P;Q) = \sum_{i = 1}^n \Big( g(p_i) - g(q_i) - g'(q_i)(p_i - q_i) \Big).
\end{align}
In this case,  $D^G(\cdot ; \cdot)$ is said to be a decomposable Bregman divergence.

\subsection{$f$-Divergences}

Morimoto \cite{Morimoto1963markov}, Csisz{\'a}r\cite{Csiszar1967information}, and Ali and Silvey\cite{Ali--Silvey1966} independently introduced the notion of  $f$-divergences, which take the form
\begin{align}
D_f(P;Q) \triangleq \sum_{i = 1}^n q_i f \left( \frac{p_i}{q_i} \right),
\end{align}
where $f$ is a convex function satisfying $f(1) = 0$. By Jensen's inequality, the convexity of $f$ ensures that $f$-divergences are always nonnegative:
\begin{equation}
D_f(P;Q) \geq f \left( \sum_{i = 1}^n  q_i \frac{p_i}{q_i} \right) = 0,
\end{equation}
and therefore are consistent with our general definition of a divergence.
Well-known examples of $f$-divergences include the Kullback--Leibler divergence, Hellinger distance, $\chi^2$-divergence, and total variation distance. From their definition, it is immediate that all $f$-divergences are decomposable. Many important properties of $f$-divergences can  be found in Basseville's survey \cite{Basseville2013divergence} and references therein.

\subsection{Kullback-Leibler-type $f$-distance measures}
A {Kullback-Leibler-type $f$-distance measure}  (or, KL-type $f$-distance measure) \cite{Kannappan1998} takes the form
\begin{align}
L(P;Q) = \sum_{k = 1}^n p_k \Big( f(q_k) - f(p_k) \Big) \geq 0. \label{KLtypeDefn}
\end{align}
If a particular divergence $L(P;Q)$ is defined by \eqref{KLtypeDefn} for a given $f$, we say that $f$ generates $L(P;Q)$.  Theorems \ref{thm.fischer} and \ref{thm.fischern2} characterize all permissible functions $f$ which generate KL-type $f$-distance measures.  Indeed, when $n\geq 3$, any KL-type $f$-distance measure is proportional to the standard Kullback-Leibler divergence.  As with $f$-divergences, KL-type $f$-distance measures are decomposable by definition.

\section{Main Results}\label{sec:mainResults}
In this section, we address each of the questions posed in the introduction.  A subsection is devoted to each question, and all proofs of stated results can be found in the appendices. 

\subsection{\Q1: Are $f$-Divergences the unique decomposable divergences which satisfy {\sc Data Processing}?} \label{sec:Q1Answer}
Recall that a {decomposable divergence} $D$ takes the form:
\begin{align}
D(P ; Q) = \sum_{i = 1}^n \delta(p_i, q_i),
\end{align}
where $\delta(u,v): [0,1]^2 \to \bar{\bR}$ is an arbitrary bivariate function.  Theorem~1 in \cite{Pardo--Vajda1997} asserts that any decomposable divergence which satisfies {\sc Data Processing} must be an $f$-divergence.  However, the proof of \cite[Theorem 1]{Pardo--Vajda1997} only works when $n\geq 3$, a fact which  apparently went unnoticed\footnote{The propagation of this problem influences other claims in the literature (cf. \cite[Theorem 2]{Harremoes--Tishby2007}).}.  The proof of the same claim in \cite[Appendix]{Amari2009} suffers from a similar flaw and also fails when $n=2$.  Of course, knowing that  the assertion holds for $n\geq 3$, it is natural to expect that it also must hold for $n=2$.  As it turns out, this is not true.  In fact, counterexamples exist in great abundance.

To this end, take any $f$-divergence $D_f(P;Q)$ and let  $k: \bR \to \bR$ be an arbitrary nondecreasing function, such that $k(0) = 0$.  Since all  $f$-divergences satisfy {\sc Data Processing} (cf. \cite{Basseville2013divergence} and references therein), the divergence function $\tilde{D}(P;Q) \triangleq k\left(D_f(P;Q)\right)$ must also satisfy {\sc Data Processing}. It was first observed in Amari~\cite{Amari2009}. It turns out that the divergence function $\tilde{D}(P;Q)$ is also decomposable in the binary case, which follows immediately from decomposability of $f$-divergences and the following lemma, which is proved in the appendix.
\begin{lemma}\label{lemma.alwaysdec}
A divergence function $D$ on a binary alphabet  is decomposable  if and only if
\begin{align}
D((p,1-p); (q,1-q)) = D((1-p,p); (1-q,q)).
\end{align}
\end{lemma}

Therefore, if $\tilde{D}(P;Q)$ is not itself an $f$-divergence, we can conclude that $\tilde{D}(P;Q)$ constitutes a counterexample to \cite[Theorem 1]{Pardo--Vajda1997} for the binary case.  Indeed, $\tilde{D}(P;Q)$ is generally not an $f$-divergence, and a concrete example is as follows. Taking $f(x) = |x-1|$, we have
\begin{equation}
D_f(P ; Q) = \sum_{i = 1}^n |p_i - q_i|,
\end{equation}
which, in the binary case,  reduces to
\begin{equation}
D_f\Big((p,1-p);(q,1-q)\Big) = 2 |p-q|.
\end{equation}
Letting $k(x) = x^2$, we have
\begin{align}
\tilde{D}(P;Q) & = 4 (p-q)^2 \\
&  = 2 (p-q)^2 + 2 \left( (1-p) - (1-q) \right)^2 \\
& = \delta(p,q) + \delta(1-p,1-q),
\end{align}
where $\delta(p,q) = 2(p-q)^2$. Since $\tilde{D}(P;Q) = 4 (p-q)^2$ \footnote{The divergence $(p-q)^2$ on binary alphabet is called the Brier score \cite{Brier1950verification}.} is a Bregman divergence, we will see later in Theorem \ref{thm.bregfkl} that it cannot also be an $f$-divergence because it is not proportional to KL-divergence. Thus, the answer to \Q1 is negative for the case $n=2$. What is more, we emphasize that an decomposable divergence that satisfies {\sc Data Processing} on the binary alphabet needs not to be a function of an $f$-divergence. Indeed, for any two $f$-divergences $D_{f_1}(P;Q),D_{f_2}(P,Q)$, $k(D_{f_1}(P;Q), D_{f_2}(P;Q))$ is also a divergence on binary alphabet satisfying {\sc Data Processing}, if $k(\cdot,\cdot)$ is nonnegative and nondecreasing for both arguments. The fact that a divergence satisfying {\sc Data Processing} does not need to be a function of an $f$-divergence was already observed in Polyanskiy and Verd\'u \cite{Polyanskiy--Verdu2010arimoto}. 

As mentioned above,  \cite[Theorem 1]{Pardo--Vajda1997} implies the answer is affirmative when $n\geq 3$.

\subsection{\Q2: Is KL divergence the only KL-type $f$-distance measure which satisfies {\sc Data Processing}?} \label{sec:Q2Answer}

Recall from Section \ref{sec:introToDivergence} that a {KL-type $f$-distance measure}  takes the form
\begin{align}
L(P;Q) = \sum_{k = 1}^n p_k \Big( f(q_k) - f(p_k) \Big). 
\end{align}
If a particular divergence $L(P;Q)$ is defined by \eqref{KLtypeDefn} for a given $f$, we say that $f$ generates $L(P;Q)$.

As alluded to in the introduction, there is a dichotomy between  {KL-type $f$-distance measures} on binary alphabets, and those on larger alphabets.  In particular, we have the following:
\begin{theorem}\label{thm.typefdpi}
If $L(P;Q)$ is a {Kullback-Leibler-type $f$-distance measure} which satisfies {\sc Data Processing}, then
\begin{enumerate}
\item If $n\geq 3$, $L(P;Q)$ is equal to  KL divergence up to a nonnegative multiplicative factor;
\item If $n = 2$ and the function $f(x)$ that generates $L(P;Q)$ is continuously differentiable, then $f(x)$ is of the form 
\begin{align}
f(x) = \int \frac{G(x)}{x} dx, \quad \mbox{for~}x \in (0,1),
\end{align}
where $G(x)$ satisfies the following properties:
\begin{enumerate}
\item $x G(x) = (x-1)h(x)$ for $x\in (0,1/2]$ and some nonnegative, nondecreasing continuous function $h(x)$.
\item $G(x) = G(1-x)$ for $x\in[1/2,1)$.
\end{enumerate}
Conversely, any nonnegative, non-decreasing continuous function $h(x)$ generates a KL-type divergence in the manner described above which satisfies {\sc Data Processing}. 
\end{enumerate}
\end{theorem}

To illustrate the last claim of Theorem \ref{thm.typefdpi}, take for example $h(x) = x^2, x\in [0,1/2]$.  In this case, we obtain
\begin{align}
f(x) = \frac{1}{2}x^2 -x +C, ~~\forall x \in [0,1], \label{eqn:fExample}
\end{align}
where $C$ is a constant of integration.  Letting $P = (p,1-p), Q= (q,1-q)$, and plugging \eqref{eqn:fExample} into \eqref{KLtypeDefn}, we obtain the KL-type divergence
\begin{equation}
L(P;Q) = \frac{1}{2} (p-q)^2 
\end{equation}
which  satisfies  {\sc Data Processing}, but  certainly does not equal KL divergence up to a nonnegative multiplicative factor.  Thus, the answer to  \Q2 is negative.

At this point it is instructive to compare with the discussion on \Q1.  In Section \ref{sec:Q1Answer}, we showed that a divergence which is decomposable and satisfies {\sc Data Processing} is not necessarily an $f$-divergence when the alphabet is binary.  From the above example, we see that the much  stronger hypothesis -- that a divergence is a Kullback-Leibler-type $f$-distance measure which satisfies {\sc Data Processing} --  still does not necessitate an $f$-divergence in the binary setting.

\subsection{\Q3: Is KL divergence the unique Bregman divergence which satisfies {\sc Sufficiency}?}

In this section, we investigate whether  KL divergence is the unique Bregman divergence that satisfies {\sc Sufficiency}. Again, the answer to this is affirmative for $n\geq 3$, but negative in the binary case.  This is captured by the following theorem.

\begin{theorem}\label{thm.bregkl}
If $D^G(P;Q)$ is a  Bregman divergence which satisfies {\sc Sufficiency} and
\begin{enumerate}
\item $n\geq 3$, then $D^G(P;Q)$ is equal to the KL divergence up to a nonnegative multiplicative factor;
\item $n = 2$, then $D^G(P;Q)$ can be any Bregman divergence generated by a symmetric bivariate convex function $G(P)$ defined on $\Gamma_2$.   
\end{enumerate}
\end{theorem}

The first part of Theorem~\ref{thm.bregkl} is possibly surprising, since we do not assume the Bregman divergence $D^G(P;Q)$ to be decomposable a priori. In an informal remark immediately following Theorem 2 in \cite{Harremoes--Tishby2007}, a claim similar to first part of our Theorem \ref{thm.bregkl} was proposed.  However, we are the first to give a complete proof of this result, as no proof was previously known \cite{HarremoesPrivate}.

We have already seen an example of a Bregman divergence which satisfies {\sc Data Processing} (and therefore {\sc Sufficiency}) in our previous examples.  Letting $P = (p,1-p), Q= (q,1-q)$ and defining $G(P) = p^2 + (1-p)^2$ generates the Bregman divergence
\begin{align}
D^G(P;Q) = 2(p-q)^2.
\end{align}
The second part of Theorem \ref{thm.bregkl} characterizes all Bregman divergences on binary alphabets which satisfy {\sc Sufficiency} as being in precise correspondence with the set of symmetric bivariate convex functions defined on $\Gamma_2$.

It is worth mentioning that Theorem~\ref{thm.typefdpi} is closely related to Theorem~\ref{thm.bregkl} in the binary case. According to the Savage representation\cite{Gneiting--Raftery2007}, there exists a bijection between Bregman divergences and KL-type $f$-distances on the binary alphabet. Hence, Theorem~\ref{thm.typefdpi} implies that KL divergence is not the unique Bregman divergence even if we restrict it to satisfy {\sc Data Processing} on the binary alphabet. In fact, Theorem~\ref{thm.typefdpi} characterizes a wide class of Bregman divergences that satisfy {\sc Data Processing} on binary alphabet, but do not coincide with KL divergence. In contrast, Theorem~\ref{thm.bregkl} characterizes the set of Bregman divergences that satisfy {\sc Sufficiency} in the binary case.

\subsection{\Q4: Is KL divergence the unique Bregman divergence which is also an $f$-divergence?}

We conclude our investigation by asking whether  Kullback--Leibler divergence is the unique divergence which is both a  Bregman divergence and an $f$-divergence. The first result of this kind was proved in \cite{Csiszar1991} in the context of linear inverse problems requiring an alphabet size of $n\geq 5$, and is hard to extract as an independent result. This question has also been considered in \cite{Harremoes--Tishby2007, Amari2009}, in which the answer is shown to be affirmative when $n\geq 3$. Note that since any $f$-divergence satisfies {\sc sufficiency}, Theorem \ref{thm.bregkl} already implies this result for $n \geq 3$. However, we now complete the story by showing that for all $n \geq 2$, KL divergence is the unique Bregman divergence which is also an $f$-divergence.


\begin{theorem}\label{thm.bregfkl}
Suppose $D(P;Q)$ is both a  Bregman divergence and an $f$-divergence for some $n\geq 2$.  Then $D(P;Q)$ is equal to KL divergence up to a nonnegative multiplicative factor. 
\end{theorem}
\subsection{Review}

In the previous four subsections, we  investigated the four questions posed in Section~\ref{sec.intro}. We now take the opportunity to collect these results and summarize them in terms of the alphabet size $n$.

\begin{enumerate}
\item For an alphabet size of $n\geq 3$, 

\begin{enumerate}
\item Pardo and Vajda \cite{Pardo--Vajda1997} showed any decomposable divergence that satisfies {\sc Sufficiency} must be an $f$-divergence.
\item Fischer \cite{Fischer1972} showed that any KL-type $f$-distance measure must be Kullback--Leibler divergence.
\item The present paper proves that any (not necessarily decomposable) Bregman divergence that satisfies {\sc Sufficiency} must be Kullback--Leibler divergence.
\end{enumerate}
\item In contrast, for binary alphabets of size of $n= 2$, we have shown in this paper that
\begin{enumerate}
\item A decomposable divergence that satisfies {\sc Data Processing} does not need to be an $f$-divergence (Section~\ref{sec:Q1Answer}).
\item A KL-type $f$-distance measure that satisfies {\sc Data Processing} does not need to be Kullback--Leibler divergence (Theorem~\ref{thm.typefdpi}).  Moreover, a complete characterization of this class of divergences is given.
\item A Bregman divergence that satisfies {\sc Sufficiency} does not need to be Kullback--Leibler divergence (Theorem~\ref{thm.bregkl}), and this class of divergences is completely characterized. 
\item The only divergence that is both a Bregman divergence and an $f$-divergence is Kullback--Leibler divergence (Theorem~\ref{thm.bregfkl}).
\end{enumerate}
\end{enumerate}

\section{Concluding Remarks}\label{sec:Conclusion}

Motivated partly by the dichotomy between the binary- and larger-alphabet settings in the characterization of Shannon entropy using Shannon-type inequalities \cite[Chap. 4.3]{Aczel--Darczy1975}, we investigate the curious case of binary alphabet in more general scenarios. In the four problems we consider, three of them exhibit a similar dichotomy between binary and larger alphabets. Concretely, we show that $f$-divergences are not the unique decomposable divergences on binary alphabets that satisfy the data processing inequality, thereby clarifying claims that have previously appeared in \cite{Pardo--Vajda1997},\cite{Amari2009},\cite{Harremoes--Tishby2007}. We show that KL divergence is the only KL-type $f$-distance measure which satisfies the data processing inequality, only when the alphabet size is at least three. To the best of our knowledge, we are the first to demonstrate that, without assuming the decomposability condition, KL divergence is the unique Bregman divergence which is invariant to statistically sufficient transformations of the data when the alphabet size is larger than two, a property that does not hold for the binary alphabet case. Finally, we demonstrate that KL divergence is the unique Bregman divergence which is also an $f$-divergence, on any alphabet size using elementary methods, which is a claim made in the literature either in different settings \cite{Csiszar1991}, or proven only in the case $n\geq 3$ \cite{Harremoes--Tishby2007,Amari2009}.

\section*{Acknowledgment}

We would like to thank Jingbo Liu for suggesting the parametrization trick in the proof of Theorem~\ref{thm.bregfkl}. We would also like to thank Peter Harremo\"es for pointing out the seminal work by Kullback and Leibler\cite{Kullback--Leibler1951}, which supports Theorem~\ref{thm.bregkl} in demonstrating the importance of KL divergence in information theory and statistics. We thank an anonymous reviewer for a careful analysis of 
the proof of Theorem~\ref{thm.bregfkl}, allowing us to relax the assumptions on function $f$ from an earlier version of this paper. 

\appendices

\section{Proof of Lemma~\ref{lemma.alwaysdec}}

Define $h(p,q) \triangleq D( (p,1-p) ; (q,1-q))$, where $D(\cdot ; \cdot)$ is an arbitrary divergence function on the binary alphabet. We first prove the ``if" part:

 If 
$D((p,1-p);(q,1-q)) = D((1-p,p); (1-q,q))$, 
then 
$h(p,q) = h(\bar{p},\bar{q})$, where $\bar{p} = 1-p,\bar{q} = 1-q$. To this end, define 
\begin{equation}
\delta(p,q) = \frac{1}{2}h(p,q),
\end{equation}
and note that we have that
\begin{equation}
h(p,q) = \delta(p,q) +\delta(\bar{p},\bar{q}).
\end{equation}
This implies that $D((p,1-p) ; (q,1-q))$ is a decomposable divergence.

Now we show the ``only if" part: 
Suppose there exists a function $\delta(p,q):[0,1]\times [0,1]\to \bar{\bR}$, such that
\begin{equation}
h(p,q) = \delta(p,q) + \delta(\bar{p},\bar{q}).
\end{equation}
Then,
\begin{equation}
h(\bar{p},\bar{q}) = \delta(\bar{p},\bar{q}) + \delta(p,q) = h(p,q),
\end{equation}
which is equivalent to
\begin{equation}
D((p,1-p); (q,1-q)) = D((1-p,p); (1-q,q)),
\end{equation}
completing the proof.

\section{Proof of Theorem~\ref{thm.typefdpi}}
When $n\geq 3$, Theorem~\ref{thm.fischer} implies that $L(P;Q)$ is equal to the KL divergence up to a non-negative multiplicative factor. Hence it suffices to deal with the binary alphabet case. 
Since $Y \in \mathcal{X}$ is also a binary random variable, we may parametrize any (stochastic) transform between two binary alphabets by the following binary channel:
\begin{align*}
P_{Y|X}(1|1) & = \alpha \\
P_{Y|X}(2|1) & = 1-\alpha \\
P_{Y|X}(1|2) & = \beta \\
P_{Y|X}(2|2) & = 1-\beta, 
\end{align*}
where $\alpha,\beta \in [0,1]$ are parameters. 
Under the binary channel $P_{Y|X}$, if the input distribution is $P_X=(p,1-p)$, the output distribution $P_Y$ would be $(p \alpha + \beta (1-p), (1-\alpha)p + (1-\beta) (1-p))$. For notational convenience, denote $\tilde{p} = p \alpha + \beta (1-p), \tilde{q} = q \alpha + \beta (1-q)$. 
The data processing property implies that
\begin{align}
& p \left( f(q) -f(p) \right) + (1-p) \left( f(1-q) - f(1-p) \right) \nonumber \\
& \quad \geq \tilde{p} \left( f(\tilde{q}) - f(\tilde{p}) \right) + (1-\tilde{p}) \left( f(1-\tilde{q}) - f(1-\tilde{p}) \right),\label{eqn.data}
\end{align}
for all $p,q,\alpha,\beta \in [0,1]$. 
Taking $\alpha = \beta = 1$, we obtain
\begin{equation}
p \left( f(q) -f(p) \right) + (1-p) \left( f(1-q) - f(1-p) \right) \geq 0.
\end{equation}
Theorem~\ref{thm.fischern2} gives the general solution to this functional inequality. In particular, it implies that there exists a function $G(p)$ such that
\begin{equation}
f'(p) = G(p)/p,\qquad G(p)\leq 0, \qquad G(p) = G(1-p).
\end{equation}
Note that both sides of (\ref{eqn.data}) are zero when $q = p$. Since we assume $f$ is continuously differentiable, we know that there must exist a positive number $\delta>0$, such that for any $q\in [p,p+\delta)$, the derivative of LHS of (\ref{eqn.data}) with respect to $q$ is no smaller than the derivative of the RHS of (\ref{eqn.data}) with respect to $q$.
Hence, we assume $q\in [p,p+\delta)$ and take derivatives with respect to $q$ on both sides of (\ref{eqn.data}) to obtain
\begin{equation}
p f'(q) - (1-p) f'(1-q) \geq \tilde{p} (\alpha - \beta) f'(\tilde{q}) + (1-\tilde{p})(\beta - \alpha) f'(1-\tilde{q}).
\end{equation} 
Substituting $f'(p) = G(p)/p, G(p) = G(1-p)$, we find that
\begin{equation}
\frac{p-q}{q(1-q)} G(q) \geq (\alpha - \beta) G(\tilde{q}) \frac{(p-q)(\alpha-\beta)}{\tilde{q}(1-\tilde{q})}.
\end{equation}
Since we assumed that $q > p$, we have shown
\begin{equation}\label{eqn.ineq}
\frac{G(q)}{G(\tilde{q})} \geq \frac{(\alpha-\beta)^2 q (1-q)}{\tilde{q}(1-\tilde{q})}.
\end{equation}

Noting that  $p$ does not appear in (\ref{eqn.ineq}), we know (\ref{eqn.ineq}) holds for all $q,\alpha,\beta \in [0,1]$. However, the four parameters $\alpha,\beta,q,\tilde{q}$ are not all free parameters, since $\tilde{q}$ is completely determined by $q,\alpha,\beta$ through its definition $\tilde{q} = q \alpha + \beta (1-q)$. In order to eliminate this dependency, we try to eliminate $\alpha$ in (\ref{eqn.ineq}).  
Since 
\begin{equation}\label{eqn.alpha}
\alpha = \frac{\tilde{q} - \beta (1-q)}{q} \in [0,1],
\end{equation} 
we know that
\begin{equation}
\tilde{q} \in [\beta (1-q), q + \beta (1-q)],
\end{equation}
which is equivalent to the following constraint on $\beta$:
\begin{equation}
\max \left( 0, \frac{\tilde{q} - q}{1-q} \right) \leq \beta \leq \min \left( \frac{\tilde{q}}{1-q}, 1\right). 
\end{equation}
Plugging (\ref{eqn.alpha}) into (\ref{eqn.ineq}), we obtain
\begin{equation}\label{eqn.finalin}
\frac{G(q)}{G(\tilde{q})} \geq \frac{(\tilde{q}-\beta)^2 (1-q)}{q \tilde{q}(1-\tilde{q})}.
\end{equation}
 
In order to get the tightest bound in (\ref{eqn.finalin}), we need to maximize the RHS of (\ref{eqn.finalin}) with respect to $\beta$. By elementary  algebra, it follows from the symmetry of $G(x)$ with respect to $x = 1/2$ that (\ref{eqn.finalin}) can be reduced to the following inequality:
\begin{equation}\label{eqn.ineqxy}
\frac{G(x)}{G(y)} \geq \frac{y(1-x)}{x(1-y)}, \qquad 0\leq y\leq  x\leq 1/2.
\end{equation} 
Inequality~(\ref{eqn.ineqxy}) is equivalent to
\begin{equation}
G(x) \frac{x}{1-x} \leq G(y) \frac{y}{1-y}, \qquad 0\leq y\leq  x\leq 1/2, 
\end{equation}
which holds if and only if 
\begin{equation}
G(x) \frac{x}{1-x}
\end{equation}
is a non-positive non-increasing function on $[0,1/2]$. 
In other words, there exists a function $h(x), x\in [0,1/2]$, such that
\begin{equation}
G(x) = \frac{x-1}{x} h(x), 
\end{equation}
$h(x)\geq 0$, $h(x) \textrm{ non-decreasing } \forall x\in [0,1/2]$. To conclude, the data processing inequality implies that the derivative of $f(x)$ admits the following representation:
\begin{equation}\label{eqn.solf}
f'(x) = G(x)/x,
\end{equation}
where for $x\in (0,1/2]$, $G$ satisfies
\begin{align}
G(x) & = G(1-x) \\
x G(x) & = (x-1)h(x),
\end{align}
and $h(x)\geq 0$ is a non-decreasing continuous function on $(0,1/2]$. 

Conversely, for any $f$ whose derivative can be expressed in the form (\ref{eqn.solf}), we can show the data processing inequality holds. Indeed, for a function $f$ admitting representation (\ref{eqn.solf}), it suffices to show for any $\alpha,\beta\in [0,1]$, the derivative of LHS with respect to $q$ is larger than the derivative of RHS with respect to $q$ in (\ref{eqn.data}) when $q>p$, and is smaller when $q<p$. This follows as a consequence of the previous derivations.

\section{Proof of Theorem~\ref{thm.bregkl}}
Before we begin the proof of Theorem \ref{thm.bregkl}, we take the opportunity to state a symmetry lemma that will be needed.
\begin{lemma}\label{lem:BregSymm}
If the Bregman divergence $D^G(P;Q)$ generated by $G$ satisfies {\sc Sufficiency}, then there is a symmetric convex function $G_{\mathsf{s}}$ that generates the same Bregman divergence. That is,  $D^G(P;Q)=D^{G_{\mathsf{s}}}(P;Q)$ for all $P,Q\in \Gamma_n$.
\end{lemma}
\begin{IEEEproof}
Let $N=(1/n, \dots, 1/n)\in \Gamma_n$ denote the uniform distribution on $\mathcal{X}$, and consider a permutation $Y=\pi(X)$.  Since $D^G(P;Q)$ obeys {\sc Sufficiency}, it follows that 
\begin{align}
& G(P) - G(N) - \langle \nabla G(N), P - N \rangle \nonumber \\
& \quad = G(P_{\pi^{-1}}) - G(N) - \langle \nabla G(N), P_{\pi^{-1}} - N \rangle ,
\end{align}
where $P_{\pi^{-1}}(i) = p_{\pi^{-1}(i)}$ for $1\leq i \leq n$. This implies that $G_{\mathsf{s}}(P) \triangleq G(P) - \langle \nabla G(N), P \rangle$ is a symmetric convex function on $\Gamma_n$.  Since Bregman divergences are invariant to affine translations of the generating function, the claim is proved.
\end{IEEEproof}

Note that Lemma \ref{lem:BregSymm} essentially proves the Theorem's assertion for $n=2$.  Noting that the only sufficient transformation on binary alphabets is a permutation of the elements finishes the proof.  Therefore, we only need to consider the setting where $n\geq 3$.  This will be accomplished in the following series of propositions.

\begin{proposition}\label{prop:BregmanSuff}
Suppose $n\geq 3$.  If $D(P;Q)$ is a Bregman divergence on $\Gamma_n$ that satisfies {\sc Sufficiency}, then there exists a convex function $G$ that generates $D(P;Q)$ and admits the following representation:
\begin{equation}
G(P) = (p_1 + p_2) U \left( \frac{p_1}{p_1 + p_2} ;  p_4,\ldots,p_n \right) + E (p_1 + p_2 ;  p_4,\ldots,p_n), \label{FnlProp}
\end{equation}
where $P = (p_1,p_2,\ldots,p_n)\in \Gamma_n$ is an arbitrary probability vector, and $U(\cdot;p_4,\ldots,p_n), E(\cdot; p_4,\ldots,p_n)$ are two univariate functions indexed by the parameter $p_4,\ldots,p_n$. 
\end{proposition}
\begin{IEEEproof}
Take $P_{\lambda_1}^{(t)}, P_{\lambda_2}^{(t)}$ to be two probability vectors parameterized in the following way:
\begin{align}
P_{\lambda_1}^{(t)} & =(\lambda_1 t, \lambda_1 (1-t), r- \lambda_1  , p_4, \ldots ,p_n) \\
 P_{\lambda_2}^{(t)} & = (\lambda_2 t,\lambda_2  (1-t), r-\lambda_2 , p_4, \ldots,p_n),
\end{align}
where $r \triangleq 1- \sum_{i\geq 4} p_i, t\in [0,1], \lambda_1 < \lambda_2$.  Observe that
\begin{equation}
D^G(P_{X|1}; P_{X|2}) = G(P_{\lambda_1}^{(t)}) - G(P_{\lambda_2}^{(t)}) - \langle \nabla G(P_{\lambda_2}^{(t)}), P_{\lambda_1}^{(t)} - P_{\lambda_2}^{(t)} \rangle
\end{equation}
because $D^G(P;Q) = G(P) - G(Q) - \langle \nabla G(Q), P - Q \rangle$ by definition.  

Since the first two elements of $P_{\lambda_1}^{(t)}$ and $P_{\lambda_2}^{(t)}$ are proportional, the  transformation $X\mapsto Y$ defined by 
\begin{equation}
Y = \begin{cases} 1 & X \in \{1,2\} \\ X & \textrm{otherwise}\end{cases}
\end{equation}
is sufficient for $Z\in\{1,2\}$, where $P_{X|Z} \triangleq P_{\lambda_Z}^{(t)}$.  By our assumption that  $D^G(P;Q)$ satisfies {\sc Sufficiency}, we can conclude that  
\begin{align}
& G(P_{\lambda_1}^{(t)}) - G(P_{\lambda_2}^{(t)}) - \langle \nabla G(P_{\lambda_2}^{(t)}), P_{\lambda_1}^{(t)} - P_{\lambda_2}^{(t)} \rangle \nonumber \\
& \quad =  G(P_{\lambda_1}^{(1)}) - G(P_{\lambda_2}^{(1)}) - \langle \nabla G(P_{\lambda_2}^{(1)}), P_{\lambda_1}^{(1)} - P_{\lambda_2}^{(1)} \rangle\label{eqnWithG}
\end{align}
for all legitimate $\lambda_2 >\lambda_1 \geq 0$.

Fixing $p_4,p_5,\ldots,p_n$, define
\begin{equation}
R(\lambda,t; p_4,p_5,\ldots,p_n) \triangleq G(P_{\lambda}^{(t)}).
\end{equation}
For notational simplicity, we will denote $R(\lambda,t; p_4,p_5,\ldots,p_n) $ by $R(\lambda,t)$ since $p_4,p_5,\ldots,p_n$  remain fixed for the rest of the proof. Thus, \eqref{eqnWithG} becomes
\begin{align}
& R(\lambda_1,t) - R(\lambda_2,t) - \langle \nabla R(\lambda_2,t), P_{\lambda_1}^{(t)} - P_{\lambda_2}^{(t)} \rangle \nonumber \\
& \quad = R(\lambda_1,1) - R(\lambda_2,1) - \langle \nabla R(\lambda_2,1), P_{\lambda_1}^{(1)} - P_{\lambda_2}^{(1)} \rangle,
\end{align}
where we abuse notation and have written $\nabla R(\lambda,t)$ in place of $\nabla G(P_{\lambda}^{(t)})$.

For arbitrary real-valued functions $U(t), F(t)$ to be defined shortly, define $\tilde{R}(\lambda,t) = R(\lambda,t)-\lambda U(t) - F(t)$.  For all admissible $\lambda_1,\lambda_2,t$, it follows that
\begin{align}
& R(\lambda_1,t) - R(\lambda_2,t) - \langle \nabla R(\lambda_2,t), P_{\lambda_1}^{(t)} - P_{\lambda_2}^{(t)} \rangle \nonumber \\
& \quad = \tilde{R}(\lambda_1,t) - \tilde{R}(\lambda_2,t) - \langle \nabla \tilde{R}(\lambda_2,t), P_{\lambda_1}^{(t)} - P_{\lambda_2}^{(t)} \rangle, 
\end{align}
which implies that
\begin{align}
& \tilde{R}(\lambda_1,t) - \tilde{R}(\lambda_2,t) - \langle \nabla \tilde{R}(\lambda_2,t), P_{\lambda_1}^{(t)} - P_{\lambda_2}^{(t)} \rangle \nonumber \\
& \quad  = \tilde{R}(\lambda_1,1) - \tilde{R}(\lambda_2,1) - \langle \nabla \tilde{R}(\lambda_2,1), P_{\lambda_1}^{(1)} - P_{\lambda_2}^{(1)} \rangle. \label{eqn.abtoplug}
\end{align}
Fixing $\lambda_2$, we can choose the functions $U(t), F(t)$ so that $\tilde{R}(\lambda_2,t)$ satisfies
\begin{align}
\tilde{R}(\lambda_2,t) & =\tilde{R}(\lambda_2,1) \nonumber \\
 \langle \nabla \tilde{R}(\lambda_2,t), P_{\lambda_1}^{(t)} - P_{\lambda_2}^{(t)} \rangle & =   \langle \nabla \tilde{R}(\lambda_2,1), P_{\lambda_1}^{(1)} - P_{\lambda_2}^{(1)} \rangle. \label{eqn.abdesing}
\end{align}
Plugging (\ref{eqn.abdesing}) into (\ref{eqn.abtoplug}), we find that
\begin{equation}
\tilde{R}(\lambda_1,t) =  \tilde{R}(\lambda_1,1).
\end{equation}
Therefore, there must exist a function $E: [0,1] \to \bR$, such that 
\begin{equation}
\tilde{R}(\lambda,t) = E(\lambda).
\end{equation}

By definition, $R(\lambda,t) = \tilde{R}(\lambda,t) + \lambda U(t) + F(t)$.  Hence, we can conclude that there exist real-valued functions $E, U,F$ (indexed by $p_4,\ldots,p_n$) such that
\begin{equation}
R(\lambda,t) = F(t) + \lambda U(t) + E(\lambda).
\end{equation}
By definition of $R(\lambda,t)$, we have
\begin{align}
& G(p_1,p_2, p_3, p_4,\ldots,p_n) \nonumber  \\
& \quad = F \left( \frac{p_1}{p_1 + p_2};  p_4,\ldots,p_n \right) \nonumber \\
& \quad \quad + (p_1 + p_2) U \left( \frac{p_1}{p_1 + p_2} ;  p_4,\ldots,p_n \right) \nonumber \\
& \quad \quad 
+ E (p_1 + p_2 ;  p_4,\ldots,p_n), \label{genExpG}
\end{align}
which follows from expressing $\lambda,t$ in terms of $p_1,p_2$:
\begin{equation}
\lambda = p_1 + p_2, \quad t = \frac{p_1}{p_1 + p_2}. 
\end{equation}

Reparameterizing $p_1 = xa, p_2 = x (1-a), a\in [0,1], p_3 = 1- \left( \sum_{i\geq 4}p_i\right) - x$ and letting $x \downarrow 0$, it follows from from \eqref{genExpG} that
\begin{equation}\label{eqn.elimia}
\lim \limits_{x \downarrow 0} G(P) = F \left( a ;  p_4,\ldots,p_n \right) + \lim \limits_{x\downarrow 0} E(x;  p_4,\ldots,p_n) .
\end{equation}

In words, Equation~(\ref{eqn.elimia}) implies that if $F$ is not identically constant, $\lim \limits_{x \downarrow 0} G(P)$ is going to depend on how we approach the boundary point $(0,0, 1- \sum_{i\geq 4}p_i, p_4,\ldots,p_n)$. 
Since $\Gamma_n$ is a bounded closed polytope, we obtain a contradiction by recalling:
\begin{lemma}[Gale-Klee-Rockafellar \cite{Gale--Klee--Rockafellar1968}]\label{GKR}
If $S$ is boundedly polyhedral and $\phi$ is a convex function on the relative interior of $S$ which is bounded on bounded sets, then $\phi$ has a unique continuous convex extension on $S$. Moreover, every convex function on a bounded closed polytope is bounded. 
\end{lemma}

Without loss of generality, we may take $F \equiv 0$, completing the proof.
\end{IEEEproof}

Proposition \ref{prop:BregmanSuff} establishes that Bregman divergences which satisfy {\sc Sufficiency} can only be generated by convex functions $G$ which satisfy a functional equation of the form \eqref{FnlProp}.
Toward characterizing the solutions to \eqref{FnlProp}, we cite a result on the the so-called \emph{generalized fundamental equation of information theory}.

\begin{lemma}[See \cite{Kannappan--Ng1973,Maksa1982,Aczel--Ng1983}.]\label{lemma.fundamental}
Any   measurable solution of
\begin{equation}
f(x) + (1-x) g \left( \frac{y}{1-x} \right) = h(y) + (1-y) k \left( \frac{x}{1-y} \right),\label{fundamentalFnl}
\end{equation}
for $x,y\in [0,1)$ with $x+y \in [0,1]$, where $f,h: [0,1) \to \bR$ and $g,k: [0,1] \to \bR$, has the form
\begin{align}
f(x) & = a H_2(x) + b_1 x + d, \\
g(y) & = a H_2(y) + b_2 y + b_1 - b_4, \\
h(x) & = a H_2(x) + b_3 x + b_1 + b_2 - b_3 - b_4 + d, \\
k(y) & = a H_2(y) + b_4 y + b_3 - b_2,
\end{align} 
for $x\in [0,1), y\in [0,1]$, where $H_2(x) = -x \ln x -(1-x) \ln (1-x)$ is the binary Shannon entropy and $a,b_1,b_2,b_3,b_4$, and $d$ are arbitrary constants.
\end{lemma}

We remark that if  $f = g = h = k$ in Lemma~\ref{lemma.fundamental}, the corresponding functional equation is called the \emph{fundamental equation of information theory}, and has the solution $f(x)=C\cdot H_2(x)$, where $C$ is an arbitrary constant.

We now apply Lemma \ref{lemma.fundamental} to prove the following refinement of Proposition \ref{prop:BregmanSuff}.
\begin{proposition}\label{refinedProp}
Suppose $n\geq 3$.  If $D(P;Q)$ is a Bregman divergence on $\Gamma_n$ that satisfies {\sc Sufficiency}, then there exists a symmetric 
convex function $G$ that generates $D(P;Q)$ and admits the following representation:
\begin{align}
G(P) & = A(p_4,\ldots,p_n)  \left( p_1\ln p_1 + p_2\ln p_2  + p_3\ln p_3\right) \nonumber \\
& \quad +B(p_4,\ldots,p_n),
\end{align}
where $P = (p_1,p_2,\ldots,p_n)\in \Gamma_n$ is an arbitrary probability vector, and $A(p_4,\ldots,p_n), B(p_4,\ldots,p_n)$ are symmetric functions of $p_4,\ldots,p_n$. 
\end{proposition}
\begin{IEEEproof}
Taking Lemma \ref{lem:BregSymm} together with Proposition \ref{prop:BregmanSuff}, we can assume that there is a symmetric convex function $G$ that generates $D(P;Q)$ and admits the representation
\begin{equation}
G(P) = (p_1 + p_2) U \left( \frac{p_1}{p_1 + p_2} ;  p_4,\ldots,p_n \right) + E (p_1 + p_2 ;  p_4,\ldots,p_n).\label{genform}
\end{equation}
We now massage \eqref{genform} into a form to which we can apply Lemma \ref{lemma.fundamental}. For the remainder of the proof, we suppress the explicit dependence of $U \left( \cdot ;  p_4,\ldots,p_n \right)$ and $E (\cdot ;  p_4,\ldots,p_n)$ on $p_4,\ldots,p_n$, and simply write $U(\cdot)$ and $E(\cdot)$, respectively.
First, since $G(P)$ is a symmetric function, we know that if we exchange the entries $p_1$ and $p_3$ in $P$, the  value of $G(P)$ will not change. In other words, for $r = p_1 + p_2 + p_3$, we have
\begin{align}
& (r-p_3) U \left( \frac{p_1}{r-p_3} \right) + E(r-p_3)  \nonumber \\
& \quad = (r-p_1) U \left( \frac{p_3}{r-p_1} \right) + E(r-p_1).
\end{align}
Second, we define $\tilde{E}(x) \triangleq  E(r-x)$, and can now write
\begin{equation}
(r-p_3) U \left( \frac{p_1}{r-p_3} \right) + \tilde{E}(p_3)= (r-p_1) U \left( \frac{p_3}{r-p_1} \right) + \tilde{E}(p_1).
\end{equation}
Finally, we define $q_i = p_i/r$ for $i = 1,2,3$, and $h(x) = \tilde{E}(r x)/r$ to obtain
\begin{equation}
(1-q_3) U \left( \frac{q_1}{1-q_3} \right) + h (q_3) = (1-q_1) U \left( \frac{q_3 }{1-q_1}\right) + h(q_1),
\end{equation}
which has the same form as \eqref{fundamentalFnl}.
Applying Lemma~\ref{lemma.fundamental},  we find that $b_1 = b_3, b_2 = b_4$, and 
\begin{align}
h(x) &= a H_2(x) + b_1 x + d\\
U(y) &= a H_2(y) + b_2 y + b_1 - b_2.
\end{align}

By unraveling the definitions of $h(x)$ and $\tilde{E}(x)$, and recalling the symmetric relation $H_2(x)=H_2(1-x)$, we find that
\begin{equation}
E(x) = r a H_2(x/r) + b_1(r-x) +rd.
\end{equation}
Substituting the general solutions to $U(x), E(x)$ into \eqref{genform}, we have
\begin{align}
G(P)&=a (p_1+p_2)H_2\left(\frac{p_1}{p_1 + p_2}\right) + b_2 p_1 \nonumber \\
& \quad + (b_1 - b_2)(p_1+p_2) + r a H_2\left(\frac{p_1 + p_2}{r}\right) \nonumber \\
&\quad  + b_1(r-(p_1 + p_2)) +rd\\
&=a (p_1+p_2)H_2\left(\frac{p_1}{p_1 + p_2}\right)   + r a H_2\left(\frac{p_1 + p_2}{r}\right) \nonumber \\
& \quad + b_1r -b_2 p_2 +rd.\label{invPerm}
\end{align}
Since $G(P)$ is symmetric, its value must be invariant to exchanging the values $p_1\leftrightarrow p_2$.  However, \eqref{invPerm} can only be invariant to  such permutations if $b_2\equiv 0$.  Thus, we can further simplify \eqref{invPerm} and write
\begin{align}
G(P)&=a \Big( p_1\ln p_1 + p_2\ln p_2 \nonumber \\
& \quad  + (r-p_1-p_2)\ln(r-p_1-p_2) - r \ln(r)\Big) +(b_1+d)r\\
&=A(p_4,\ldots,p_n)  \left( p_1\ln p_1 + p_2\ln p_2  + p_3\ln p_3\right) \nonumber \\
& \quad +B(p_4,\ldots,p_n),
\end{align}
where $A(p_4,\ldots,p_n)$ and $B(p_4,\ldots,p_n)$ are functions of $(p_4, \ldots,p_n)$.  By performing an arbitrary permutation on $p_4,\ldots,p_n$ and noting that $p_1,p_2,p_3$ share two degrees of freedom,  we can conclude that $A(p_4,\ldots,p_n), B(p_4,\ldots,p_n)$ must be symmetric functions as desired.
\end{IEEEproof}

We are now in a position to prove Theorem \ref{thm.bregkl}.
\begin{IEEEproof}[Proof of Theorem \ref{thm.bregkl} for $n\geq 3$] Suppose   $D(P;Q)$ is a Bregman divergence that satisfies {\sc Sufficiency}.  Then, Proposition \ref{refinedProp} asserts that there must be a symmetric convex function $G(P)$ which admits the form 
\begin{align}
G(P) & =A(p_4,\ldots,p_n)  \left( p_1\ln p_1 + p_2\ln p_2  + p_3\ln p_3\right) \nonumber \\
& \quad +B(p_4,\ldots,p_n),
\end{align}
where $A$ and $B$ are symmetric functions.  By symmetry of $G(P)$, we can exchange $p_1,p_4$ to obtain the identity
\begin{align}
& A(p_4,p_5,\ldots,p_n) \left( p_1 \ln p_1 + p_2 \ln p_2 + p_3\ln p_3 \right) \nonumber \\
& \quad + B(p_4,p_5,\ldots,p_n) \nonumber \\
&\  = A(p_1,p_5, \ldots,p_n) \left( p_4 \ln p_4 + p_2 \ln p_2 + p_3\ln p_3 \right) \nonumber \\
& \quad + B(p_1,p_5,\ldots,p_n).
\end{align}
Comparing the coefficients for $p_2\ln p_2$, it follows that $A$ must satisfy   
\begin{equation}
A(p_4,p_5,\ldots,p_n) = A(p_1,p_5, \ldots,p_n).\label{swapCoords}
\end{equation}
However, since $A$ is a symmetric function, \eqref{swapCoords} implies that $A$ is a constant. Defining the constant $a\triangleq A$, we now have
\begin{align}
& a \left( p_1 \ln p_1 + p_2 \ln p_2 + p_3\ln p_3 \right) + B(p_4,p_5,\ldots,p_n)\nonumber \\
& \quad = a\left( p_4 \ln p_4 + p_2 \ln p_2 + p_3\ln p_3 \right) + B(p_1,p_5,\ldots,p_n),
\end{align}
which is equivalent to
\begin{equation}\label{eqn.dede}
a p_1 \ln p_1 - a p_4 \ln p_4 = B(p_1,p_5,\ldots,p_n) - B(p_4,p_5,\ldots,p_n).
\end{equation}

Taking partial derivatives with respect to $p_1$ on both sides of (\ref{eqn.dede}), we obtain
\begin{equation}
a (\ln p_1 + 1) = \frac{\partial}{\partial p_1} B(p_1,p_5,\ldots,p_n),
\end{equation}
which implies that there exists a function $f(p_5,\ldots,p_n)$ such that
\begin{equation}
B(p_1,p_5,\ldots,p_n) = a p_1 \ln p_1 + f(p_5,\ldots,p_n).
\end{equation}
Recalling the symmetry of $B$, we can conclude that
\begin{equation}
B(p_4,\ldots,p_n) = \sum_{i\geq 4} a p_i \ln p_i + c,
\end{equation}
where $c$ is a constant.  To summarize, we have shown that 
\begin{equation}
G(P) = a\sum_{i = 1}^n p_i \ln p_i + c.
\end{equation}
To guarantee that $G(P)$ is convex, we must have $a\geq0$.   Since $G(P) = a\sum_{i = 1}^n p_i \ln p_i+C$ generates a Bregman divergence which is a positive multiple of KL divergence, the theorem is proved.
\end{IEEEproof}

\section{Proof of Theorem~\ref{thm.bregfkl}}

Setting $p_i = q_i = 0, i \geq 3$, and denoting $p_1$ by $p$, $q_1$ by $q$, $G(p,1-p,0,\ldots,0)$ by $h(p)$, we have
\begin{equation}\label{eqn.jingbo}
h(p) - h(q) - h'(q)(p-q) = qf \left( \frac{p}{q}\right) + (1-q)f \left( \frac{1-p}{1-q} \right).
\end{equation}

Setting $p = q$, we find  that $f(1) = 0$. The function $f$ is assumed to be convex so it has derivatives from left and right. Taking derivatives with respect to $p$ on both sides of (\ref{eqn.jingbo}), we have
\begin{equation}
h'(p) - h'(q) = f_+' \left( \frac{p}{q} \right) - f_-' \left( \frac{1-p}{1-q} \right).
\end{equation}

Taking $x = p/q$, we have
\begin{equation}
h'(xq) - h'(q)  = f_+'(x) -f_-'\left( \frac{1-xq}{1-q} \right).
\end{equation}
Assume $x>1$. Then, upon letting $q \downarrow 0$, yields
\begin{equation}
\lim \limits_{q \downarrow 0} \left(  h'(xq) - h'(q) \right) = f_+'(x) - f_-'(1).
\end{equation}
Here we have used the fact that the left derivative of a convex function is left continuous. 

We now have
\begin{align}
f_+'(x) & = f_-'(1) + \lim_{q\downarrow 0} (h'(xq) - h'(q)) \\
& = f_-'(1) + \lim_{q\downarrow 0} \left( \sum_{i = 1}^n h'(x^{\frac{n-i+1}{n}}q) - h'(x^{\frac{n-i}{n}} q) \right) \\
& = f_-'(1) + \sum_{i = 1}^n \lim_{q\downarrow 0} (h'(x^{\frac{n-i+1}{n}}q) - h'(x^{\frac{n-i}{n}} q)) \\
& = f_-'(1) + \sum_{i = 1}^n \left( f_+'(x^{1/n}) -f_-'(1) \right) \\
& = n  f_+'(x^{1/n}) - (n-1)f_-'(1).
\end{align}

Hence we get
\begin{equation}
f_+'(x) - f_-'(1) = n \left( f_+'(x^{1/n}) - f_-'(1) \right). 
\end{equation}

Taking $x = 1$, we obtain $f_-'(1) = f_+'(1)$, so that we may simply write $f'(1)$. Hence,
\begin{equation}
f_+'(x) - f'(1) = n \left( f_+'(x^{1/n}) - f'(1) \right).
\end{equation}

Introduce the increasing function $g(t) = f_+'(e^t) - f'(1)$. Then we have
\begin{equation}
g(t) = n g(t/n),\text{ for any }n\in \mathbb{N}_+.
\end{equation}

It further implies that $g(t) = t g(1)$, for all $t \in \mathbb{Q}$. Considering the fact that $g(\cdot)$ is increasing, we know $g(\cdot)$ must be a linear function, which means that $g(t) = a_+ t, t>0$ for some constant $a_+$. Therefore $f_+'(x) = a_+ \ln x + f'(1)$. By the integral relation of a convex function and its right derivative, using the fact that $f(1) = 0$, we get
\begin{equation}
f(x) = a_+ \cdot \left( x \ln x - x + 1 \right) + f'(1)(x-1),\quad x\geq 1.
\end{equation}
Similarly, there exists a constant $a_-$ such that
\begin{equation}
f(x) = a_- \cdot \left( x \ln x - x + 1 \right) + f'(1)(x-1),\quad x< 1.
\end{equation}

Plugging the expression of $f$ into (\ref{eqn.jingbo}), we obtain $a_+ = a_-$. Substituting this general solution into the definition of an $f$-divergence, we obtain
\begin{equation}
\sum q_i f \left(\frac{p_i}{q_i}\right) = f''(1) D_{\mathrm{KL}}(P \| Q),
\end{equation}
which finishes the proof.

\bibliographystyle{IEEEtran}
\bibliography{di}

\begin{IEEEbiographynophoto}{Jiantao Jiao}
(S'13) received the B.Eng. degree with the highest honor in Electronic Engineering from Tsinghua University, Beijing, China, in 2012. He is currently working towards the Ph.D. degree in the Department of Electrical Engineering, Stanford University. He is a recipient of the Stanford Graduate Fellowship (SGF). His research interests include information theory and statistical signal processing, with applications in communication, control,
computation, networking, data compression, and learning. 
\end{IEEEbiographynophoto}

\begin{IEEEbiographynophoto}{Thomas A. Courtade}
(S'06-M'13) is an Assistant Professor in the Department of Electrical
Engineering and Computer Sciences at the University of California, Berkeley.
Prior to joining UC Berkeley in 2014, he was a postdoctoral fellow supported
by the NSF Center for Science of Information. He received his Ph.D. and M.S.
degrees from UCLA in 2012 and 2008, respectively, and he graduated summa
cum laude with a B.Sc. in Electrical Engineering from Michigan Technological
University in 2007.

His honors include a Distinguished Ph.D. Dissertation award and an
Excellence in Teaching award from the UCLA Department of Electrical
Engineering, and a Jack Keil Wolf Student Paper Award for the 2012
International Symposium on Information Theory. 
\end{IEEEbiographynophoto}

\begin{IEEEbiographynophoto}{Albert No}
(S'12) is a Ph.D. candidate in the Department of Electrical Engineering at Stanford University, under the supervision of Prof. Tsachy Weissman. His research interests include relations between information and estimation theory, lossy compression, joint source-channel coding, and their applications. Albert received a Bachelor's degree in both Electrical Engineering and Mathematics from Seoul National University, in 2009, and a Masters degree in Electrical Engineering from Stanford University in 2012.
\end{IEEEbiographynophoto}

\begin{IEEEbiographynophoto}{Kartik Venkat}
(S'12) is currently a Ph.D. candidate in the Department of Electrical Engineering at Stanford University. His research interests include the interplay between information theory and statistical estimation, along with their applications in other disciplines such as wireless networks, systems biology, and quantitative finance.

Kartik received a Bachelors degree in Electrical Engineering from the Indian Institute of Technology, Kanpur in 2010, and a Masters degree in Electrical Engineering from Stanford University in 2012. His honors include a Stanford Graduate Fellowship for Engineering and Sciences, the Numerical Technologies Founders Prize, and a Jack Keil Wolf ISIT Student Paper Award at the 2012 International Symposium on Information Theory. 
\end{IEEEbiographynophoto}

\begin{IEEEbiographynophoto}{Tsachy Weissman}
(S'99-M'02-SM'07-F'13) graduated summa cum laude with a
B.Sc. in electrical engineering from the Technion in 1997, and earned
his Ph.D. at the same place in 2001. He then worked at Hewlett-Packard
Laboratories with the information theory group until 2003, when he joined
Stanford University, where he is Associate Professor of Electrical
Engineering and incumbent of the
STMicroelectronics chair in the School of Engineering.
He has spent leaves at the Technion, and at ETH Zurich.

Tsachy's research is focused on information theory, statistical signal
processing, the interplay between them, and their applications.

He is recipient of several best paper awards, and prizes for excellence in research.

He currently serves on the editorial boards of the \textsc{IEEE Transactions on Information Theory} and Foundations and Trends in Communications and
Information Theory.
\end{IEEEbiographynophoto}

\end{document}